\documentclass[preprintnumbers,superscriptaddress,byrevtex]{revtex4}
\usepackage{amsmath,amsfonts,amssymb,amscd,amsxtra,amsthm}
\usepackage{graphicx}
\usepackage{epstopdf}
\usepackage{bm}
\usepackage{stackengine}
\usepackage{feynmp}
\usepackage{feynmp-auto}
\usepackage{amsmath,amsfonts,amssymb,amscd,amsxtra,amsthm}
\begin{document}
\preprint{PKNU-NuHaTh-2022-01}
\title{A study of the elastic $\pi N$ scattering at finite baryon density}
\author{Hyeon-dong Han}
\email[E-mail: ]{monlistor@pukyong.ac.kr}
\affiliation{Department of Physics, Pukyong National University (PKNU), Busan 48513, Korea}
\affiliation{Center for Extreme Nuclear Matters (CENuM), Korea University, Seoul 02841, Korea}
\author{Parada T.~P.~Hutauruk}
\email[E-mail: ]{phutauruk@gmail.com}
 \affiliation{Department of Physics, Pukyong National University (PKNU), Busan 48513, Korea}
\author{Seung-il Nam}
\email[E-mail: ]{sinam@pknu.ac.kr}
\affiliation{Department of Physics, Pukyong National University (PKNU), Busan 48513, Korea}
\affiliation{Center for Extreme Nuclear Matters (CENuM), Korea University, Seoul 02841, Korea}
\date{\today}
\begin{abstract}
We investigate the elastic $\pi N$ scattering for the $I=3/2$ channel dominated by the $\Delta(1232)$ resonance at finite baryon density $\rho_B$ for the symmetric nuclear matter, employing the effective Lagrangian approach at the tree-level Born approximation. The quark-meson coupling (QMC) model and linear-density approximation are employed to describe the medium modifications for the in-medium baryon properties, such as the nucleon and $\Delta$ masses, and $\Delta$ full decay width. We first reproduce the experimental data in vacuum by fitting the model parameters, then analyze various physical observables including total and differential cross sections, proton-spin asymmetry in medium. It turns out that the $\Delta$-resonance contribution becomes diminished with respect to $\rho_B$, due to the weaker $\pi N$-interaction strength in medium. Beyond the $\Delta$-resonance region $\sqrt{s}\gtrsim1.5$ GeV, the cross sections are almost independent of the density. The present results will be useful to understand the role of the $\Delta$ resonance in the neutron-star equation of state (EoS) as well as in relativistic heavy-ion collision experiments. 
\end{abstract}
\maketitle
\section{Introduction}
Hadrons composed of the confined quarks that interact with the gluons in terms of quantum chromodynamics (QCD) are the generic degrees of freedom to understand the strong interactions in reality. Hadron properties such as the mass and decay width are considered to be modified in medium, due to the partial restoration of the spontaneous breakdown of chiral symmetry (SBCS), which is one of the most important ingredients to understand the low-energy non-perturbative QCD~\cite{Brown:1995qt,Hatsuda:1998vb,Rapp:1999ej}. Those modifications are extremely crucial to study the neutron star (NS) and relativistic heavy-ion collision (RHIC) experiments with finite baryon density $\rho_B$ for instance. Among the various hadrons in medium, the appearance of the $\Delta(1232,3/2^+)$ resonance becomes crucial inside NS, since the energies in the core of NS is far more sufficient to create the resonance, which is heavier than  neutron mass~\cite{Raduta:2021xiz,Dexheimer:2021sxs,Schurhoff:2010ph,Sen:2018tms,Sahoo:2018xeu,Malfatti:2020onm,Sen:2019kxt,Zhu:2016mtc}. The stability of $\Delta$-rich matter was also taken into account, using the covariant density functional theory~\cite{Raduta:2021xiz}. The effects of  the $\Delta$ resonance to the NS properties, such as the mass-radii relations were investigated in Refs.~\cite{Schurhoff:2010ph,Sen:2019kxt,Sahoo:2018xeu}, where the various QCD properties were also taken into account~\cite{Sen:2018tms,Malfatti:2020onm}. The external magnetic field induced to NS with the $\Delta$ resonance were investigated in Ref.~\cite{Dexheimer:2021sxs}. In Ref.~\cite{Zhu:2016mtc}, the authors studied the $\Delta$-resonance effects in density-dependent relativistic Hartree-Fock theory for NS. It is also worth noting that, as in Refs.~\cite{vanHees:2004sv,Rodriguez-Sanchez:2020hfh},  the medium modification in the RHIC experiments was scrutinized. Via the Dyson-Schwinger method, the $\Delta$-resonance self energy and $\pi N$ cross section in medium was studied in Ref.~\cite{Ghosh:2016hln}. In Ref.~\cite{Cui:2020fhr}, the authors studied the $\Delta$-resonance decays in isospin asymmetric nuclear matter by using the one-boson-exchange model.

Taking into account those progresses for the in-medium modification of the  $\Delta$-resonance properties, in the present work, we study the $\Delta$-resonance production through the elastic $\pi N$ scattering for the $I=3/2$ channel at finite baryon densities ($\rho_B\ne0$). Here, we assume that the incident and scattered pions are not affected by the density, since they, as the Nambu-Goldstone (NG) bosons, are moving fast and hard to interact with the medium. Moreover, the target nucleon is set to be surrounded by other nucleons uniformly. Note that the $I=3/2$ channel $(\pi^+p\to\pi^+ p)$ can eliminate the effects of the nucleon-resonance contributions in the $s$ channel, resulting in clear signals from the $\Delta$ resonance by reducing theoretical uncertainties as well. The scattering observables are computed using the effective Lagrangian method at the tree-level Born approximation in a full relativistic manner. Phenomenological form factors are taken into account for the spatial extensions of the hadrons involved. We employ various experimental and theoretical information to determine the model parameters in vacuum ($\rho_B=0$), such as the coupling constants, full decay width, and cut-off masses for the form factors. In order to consider the medium modifications of the hadrons, we make use of the quark-meson coupling (QMC) model for the symmetric nuclear matter (SNM)~\cite{Hutauruk:2018qku,Hutauruk:2019ipp,Guichon:2018uew}. In this model, the in-medium hadron masses ($M^*_{N,\Delta}$) are obtained from variation calculations using the effective Hamiltonian from the quark-meson interaction Lagrangians with the confinement bag pressure. Since we are interested in SNM, the scalar $\sigma$-meson is only taken into account. We employ the linear-density approximation for the full decay width for the $\Delta$-resonance in medium $\Gamma^*_\Delta$~\cite{Larionov:2003av,Hirata:1978wp,Oset:1987re} which also depends on energies.

As a result, it turns out that the experimental data of the total and differential cross sections (TCS and DCS) for the elastic $\pi^+p$ scattering in vacuum are qualitatively well reproduced by fitting the model parameters appropriately, showing the $\Delta$-resonance domination. As for the baryon mass modifications with respect to the baryon density via the QMC model, it turns out that the in-medium mass $M^*_{N,\Delta}$ decrease by about $15\sim20\%$ at $\rho_B\approx\rho_0$ in comparison to their vacuum values, whereas $\rho_0$ denotes the nuclear normal density. On the contrary, $\Gamma^*_\Delta$ increases as a function of $\rho_B$, indicating the interaction strength between the pion and nucleon is reduced at finite density. Using these in-medium mass and width, it is shown that TCS becomes smaller, and the $\Delta$-resonance peak gets diminished and wider obviously with respect to $\rho_B$. The peak position moves gradually to the higher-energy region as $\rho_B$ increases. In contrast, it is interesting to note that the background (BKG) contributions, except for the $\Delta$ resonance in the $s$ channel, are insensitive to the density and hardly increase with respect to $\rho_B$. The angular dependence from DCS is almost produced by the $\Delta$ resonance and the strength of DCS decreases as a function of $\rho_B$ as expected from TCS. However, as the energy increases beyond $\sqrt{s}\approx1.5$ GeV, the $\Delta$-resonance contribution is reduced and DCS almost remains the same with that for vacuum. We also compute the proton-spin asymmetry ($P$) between the target and recoil proton spin states. Hence, as for the $\Delta$-resonance region, $P$ does not change much due to the baryon density, and vice versa for the higher-energy region. We verify that the forward-scattering differential cross section $d\sigma/dt$ shows similar tendencies to those of DCS. 

This paper is organized as follows. In Sec.~II, we briefly introduce the effective Lagrangian approach and the QMC model for SNM. The numerical results and relevant discussions are given in Sec.~III. The final Section is devoted to the summary and future perspectives. 

\section{Theoretical Framework}
\subsection{Effective Lagrangians for the elastic scattering process}
\begin{figure*}[t]
  \centering
  \includegraphics[width=0.9\textwidth]{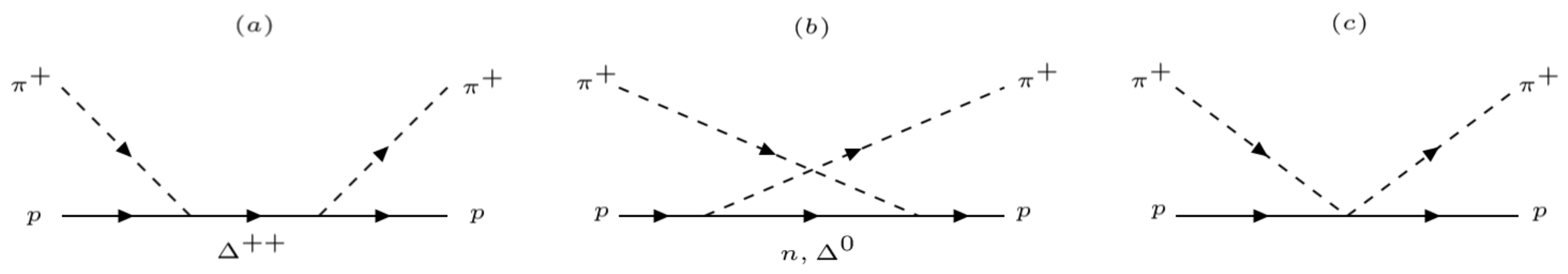}
  \caption{ \label{fig1a}(Color online) Relevant Feynman diagrams contribute to the $\pi^+$-$p$ elastic scattering for (a) the $\Delta^{++}$ pole diagram in the $s$ channel, (b) the neutron and $\Delta^0$ intermediate diagrams in the $u$ channel, and (c) Weinberg-Tomozawa (WT) contact interaction. The solid and dashed lines represent the baryon (nucleon and $\Delta$) and pion, respectively.}
\end{figure*}
In this section we briefly present the effective Lagrangian approach for the elastic scattering process. At the tree-level Born approximation, the relevant Feynman diagrams are depicted in Fig.~\ref{fig1a}: (a) the $\Delta^{++}$ pole diagram in the $s$ channel, (b) the neutron and $\Delta^0$ intermediate-state diagrams in the $u$ channel, and (c) Weinberg-Tomozawa (WT) contact interaction. The solid and dashed lines represent the baryon (nucleon and $\Delta$) and pion, respectively. For the interaction vertices, we introduce the effective Lagrangians for the corresponding channels. The effective Lagrangian for $\pi N \Delta$ vertex for the diagrams (a) and (b) in Fig.~\ref{fig1a} can be given by in terms of the Rarita-Schwinger field formalism~\cite{Gasparyan:2003fp}
\begin{eqnarray}
\label{eq1}
\mathcal{L}_{\pi N \Delta}=\frac{f_{\pi N\Delta}}{M_\pi}\bar{\Delta}^{\mu}\partial_\mu (\bm{\sigma}\cdot\bm{\pi})N + \mathrm{h.c}.
\end{eqnarray}
Thus, the corresponding scattering amplitudes are calculated as follows:
\begin{eqnarray}
\label{eq2} 
i\mathcal{M}_s^{\Delta^{++}} = -\frac{f_{\pi N\Delta}^2}{M_\pi ^2}\bar{u}(p')k'_\mu G^{\mu\nu}(p+k)k_\nu u(p),\,\,\,
i\mathcal{M}_u^{\Delta^{0}}= -\frac{f_{\pi N\Delta}^2}{3M_\pi ^2}\bar{u}(p')k_\mu G^{\mu\nu}(p-k')k'_\nu u(p),
\end{eqnarray}
with the $f_{\pi N\Delta}$ is the coupling constant of the $\pi N\Delta$. The factor of $\frac{1}{3}$ in the $u$-channel amplitude in Eq.~(\ref{eq2}) comes from the isospin factor. The $\Delta$ baryon propagator $G^{\mu\nu}(q)$ in Eq.~(\ref{eq2}) can be defined by
\begin{eqnarray}
  \label{eq3}
  G^{\mu\nu}(q) = i\frac{(\rlap{/}{q}+M_\Delta)}{q^2-M^2_\Delta-iM_\Delta\Gamma_\Delta}
 \left[-g^{\mu\nu}+\frac{1}{3}\gamma^\mu\gamma^\nu+\frac{2q^\mu q^\nu}{3M^2_\Delta}-\frac{q^\mu\gamma^\nu-q^\nu\gamma^\mu}{3M_\Delta}\right].
\end{eqnarray}

The effective Lagrangian for the $\pi NN$ interaction is then given by~\cite{Gasparyan:2003fp}
\begin{eqnarray}
  \label{eq4}
  \mathcal{L}_{\pi NN} = -\frac{f_{\pi NN} \mathcal{I}_{\pi N}}{M_\pi}\bar{N}\gamma_5\rlap{/}{\partial}(\bm{\sigma}\cdot\bm{\pi}) N+\mathrm{h.c.},
\end{eqnarray}
where $f_{\pi NN}$, $\mathcal{I}_{\pi N}$, and $M_\pi$ are respectively the pseudo-vector $\pi NN$ coupling constant (dimensionless), isospin factor of the nucleon and pion, and pion mass, whereas the $N$ and $\pi$ denote the nucleon and pion fields, respectively. Using the effective Lagrangian in Eq.~(\ref{eq4}), we straightforwardly calculate the scattering amplitude for the Feynman diagram (b) in Fig.~\ref{fig1a}, resulting in
\begin{eqnarray}
  \label{eq5}
  i\mathcal{M}_u^n = -i\frac{2}{3} \frac{f^2 _{\pi NN}}{M_\pi ^2}\bar{u}(p')\gamma_5\rlap{/}{k}\frac{\rlap{/}{p}-\rlap{/}{k}'+M_N}{(p-k')^2 -M_N^2}\gamma_5\rlap{/}{k}'u(p),
\end{eqnarray}
where $\mathcal{I}_{\pi N}=2/3$. $M_N$, $p$ and $k$ are the nucleon mass in vacuum, initial four-momenta for the nucleon and  pion, respectively, while $p'$ and $k'$ stand for the final four-momenta for them. 

The effective Lagrangian for the Weinberg-Tomozawa (WT) contact interaction for the diagram (c) in Fig.~\ref{fig1a} reads~\cite{Weinberg:1966kf,Tomozawa:1966jm,Hyodo:2011ur}
\begin{eqnarray}
  \label{eq6}
  \mathcal{L}_{\mathrm{WT}} &=& i\frac{C_{\pi N}}{4f^2_\pi}\bar{N}\left[\pi(\rlap{/}{\partial}\pi^\dagger)-(\rlap{/}{\partial}\pi)\pi^\dagger\right]N.
\end{eqnarray}
Similarly, we calculate the scattering amplitude for the contact term, using the effective Lagrangian in Eq.~(\ref{eq6}) as follows:
\begin{eqnarray}
  \label{eq7}
  i\mathcal{M}_{\mathrm{WT}} &=& -i\frac{C_{\pi N}}{4f^2_\pi}\bar{u}(p')(\rlap{/}{k}+\rlap{/}{k}')u(p).
\end{eqnarray}
The pion decay constant $f_\pi = 93.2$ MeV is used in this work and $C_{\pi N}=-1$ is for the isospin sate of $I=\frac{3}{2}$~\cite{Sun:2019nyo}.

Then, we obtain the spin-averaged differential cross section summing all the obtained scattering amplitudes in Eqs.~(\ref{eq2}), (\ref{eq4}), and (\ref{eq7}):
\begin{eqnarray}
  \label{eq8}
  \left(\frac{\partial\sigma}{\partial\Omega}\right)_{\pi^+p} &=& \frac{1}{64\pi^{2}s}\frac{|\textbf{p}|}{|\textbf{k}|}
  \frac{1}{2}\underset{\mathrm{spin}}{\sum}\left|\mathcal{M}^\mathrm{total}_{\pi^+p}\right|^2,
\end{eqnarray}
where $\textbf{k}$ and $\textbf{p}$ are the initial three-momenta for the $\pi^+$ and $p$, respectively. The total amplitude with the phenomenological form factors is written as 
\begin{eqnarray}
  \label{eq9}
\mathcal{M}^\mathrm{total}_{\pi^+p}= i\mathcal{M}_s^{\Delta^{++}}F_s^{\Delta^{++}}
  +i\mathcal{M}_u^{\Delta^{0}}F_u^{\Delta^{0}}
 +i\mathcal{M}_u^nF_u^n
 +i\mathcal{M}_{\mathrm{WT}}F_{\mathrm{WT}}, 
\end{eqnarray}
and the form factors for each channel are defined by
\begin{eqnarray}
  \label{eq9}
  F^h_{x}(x) &=& \frac{\Lambda^{4}}{\Lambda^{4}+(x-M^{2}_h)^{2}},
  \end{eqnarray}
where $x$ and $h$ denote the Mandelstam variable and corresponding hadrons, respectively. As for the WT contact amplitude, we assumed that $F_\mathrm{WT}=F^\rho_t$ for brevity, since the WT reproduces the exchange of the vector meson $\sim\rho(770)$  in the low-energy limit~\cite{Inoue:2001ip}. In this work, we set $\Lambda$ to be 800 MeV except for the WT channel, which has $\Lambda=450$ MeV, to reproduce the data as will be shown in Section III.

\subsection{Nuclear medium from the QMC model}
In this section, we present the calculation of the effective $M_N^*$ and $M_\Delta^*$ masses at finite density via the QMC model for calculating the in-medium cross sections. Besides the change of baryon masses at finite density, the momenta of the nucleon and delta baryon are expected to be changed in nuclear medium, due to the effect of the vector potential. However, in this work, since we study the in-medium effects in the nucleon rest frame, the vector potential can be ignored for brevity. In the QMC model, the nuclear medium effects come from the self-consistent exchange of the scalar $\sigma$ and vector $\omega$ meson fields which directly coupled to the confined quark in baryon~\cite{Hutauruk:2018qku,Hutauruk:2019ipp,Guichon:2018uew}. The QMC effective Lagrangian in the symmetric nuclear matter (SNM) is expressed as~\cite{Hutauruk:2018qku,Hutauruk:2019ipp}
\begin{eqnarray}
  \label{eq11}
  \mathcal{L}_\mathrm{QMC} &=& \sum_{B= N,\Delta_\nu} \bar{B} [i\gamma \cdot M_B^*(\sigma) - g_\omega \omega^\mu \gamma_\mu]B+\mathcal{L}_{M},
\end{eqnarray}
where the $g_\omega$ is the $\omega$-$N$ coupling constant, $\psi_B$ is the baryon field, and the effective baryon mass $M_B^* (\sigma)$ is given by
\begin{eqnarray}
  M_N^* (\sigma) \equiv M_N - g_\sigma (\sigma) \sigma,\,\,\,
  M_\Delta^* (\sigma) \equiv M_\Delta - g_\sigma (\sigma) \sigma,
\end{eqnarray}
where $M_N$, $M_\Delta$, and $g_\sigma (\sigma)$ are respectively the nucleon and delta masses in vacuum, and the $\sigma$-$N$ coupling constant. Thus, the free meson Lagrangian density in Eq.~(\ref{eq11}) is given by
\begin{eqnarray}
  \label{eq12}
  \mathcal{L}_M = \frac{1}{2} (\partial_\mu\sigma \partial^\mu \sigma - m_\sigma \sigma^2) - \frac{1}{2} \partial_\mu \omega_\nu(\partial^\mu \omega^\nu - \partial^\nu \omega^\mu)
 +\frac{1}{2} m_\omega^2 \omega^\mu \omega_\mu.
\end{eqnarray}
The nucleon Fermi momentum $k_F$, baryon density $\rho_B$, and scalar density $\rho_s$ in SNM at the mean-field approximation are given by
\begin{eqnarray}
  \rho_B = \frac{4}{(2\pi)^3} \int d\mathbf{k} \theta(k_F - |\mathbf{k}|) = \frac{2k_F^3}{3\pi^2},\,\,\,
  \rho_s = \frac{4}{(2\pi)^3} \int d\mathbf{k} \theta (k_F - |\mathbf{k}|) \frac{M_N^* (\sigma)}{\sqrt{M_N^{*2}(\sigma) + \mathbf{k}^2}},
\end{eqnarray}
where the Fermi momenta of the proton and neutron $k_F^{(p,n)}$ are obtained from the $\rho_p$ and $\rho_n$, respectively. In the QMC model, the Dirac equations for the light quarks ($q = u,d$) in the assumption of the non-overlapping MIT bags with the collection of nucleons treats as the nuclear matter are defined by
\begin{eqnarray}
  && \left[ i \gamma \cdot \partial_{x} - \left( m_q - V_{\sigma}^{q} \right) \mp \gamma^{0} \left( 
    V_{\omega}^{q} + \frac{1}{2} V_{\rho}^{q} \right) \right] \left( \begin{array}{c} \psi_u(x)  \\ 
    \psi_{\bar{u}}(x) \\ \end{array} \right) = 0, \nonumber \\
  && \left[ i \gamma \cdot \partial_{x} - \left( m_q - V_{\sigma}^{q} \right) \mp \gamma^{0} \left( 
    V_{\omega}^{q} - \frac{1}{2} V_{\rho}^{q} \right) \right] \left( \begin{array}{c} \psi_d(x)  \\ 
    \psi_{\bar{d}}(x) \\ \end{array} \right) = 0,
  \nonumber \\
  \label{eqintro5}
\end{eqnarray}
where the effective quark mass $m_q^{*}$ is defined by
\begin{align}
  \label{eqintro5a}
  m_q^{*} & \equiv m_q - V_{\sigma}^{q}.
\end{align}
Here, $m_q$ is the light-quark current mass ($q=u,d$) and $V_\sigma^{q}$ the scalar potential. In SNM with the Hartree approximation, the isospin dependent $\rho$-meson mean field yields $V_\rho^{q} =$ 0 in Eq.~(\ref{eqintro5}). Thus, the scalar- and vector-mean field potentials in nuclear matter are defined as
\begin{equation}
  \label{eq:potqqmc}
  V_{\sigma}^{q}  \equiv g_{\sigma}^{q} \sigma = g_{\sigma}^{q} \langle\sigma\rangle, \qquad
  V_{\omega}^{q} \equiv g_{\omega}^{q} \omega = g_{\omega}^{q} \delta^{\mu 0} \langle\omega^{\mu}\rangle.
\end{equation}

The in-medium bag radius of the hadron $R_h^{*}$ is calculated by considering the hadron mass stability condition against the variation of the bag radius, and the eigen-energies in units of $1/R_h^{*}$ are
\begin{equation}
  \label{eq:pionmed9}
  \left( \begin{array}{c}
    \epsilon_u \\
    \epsilon_{\bar{u}}
  \end{array} \right)
  = \Omega_q^* \pm R_h^* \left(  V^q_\omega + \frac{1}{2} V^q_\rho \right),\,\,\,
  \left( \begin{array}{c} \epsilon_d \\
    \epsilon_{\bar{d}}
  \end{array} \right)
 = \Omega_q^* \pm R_h^* \left(
  V^q_\omega
  - \frac{1}{2} V^q_\rho \right).
  \end{equation}
The effective hadron mass in nuclear medium $m_h^{*}$ is calculated by 
\begin{equation}
m_h^{*} = \sum_{j = q, \bar{q}} \frac{n_j \Omega_j^{*} -z_h^{}}{R^{*}_h} + \frac{4}{3} \pi  R_h^{* 3} B,\,\,\,
\Omega^{*}_q = \Omega^{*}_{\bar{q}} = \sqrt{x_q^2 + \left(R_h^{*} m_q^{*} \right)^2},
\end{equation}
while the in-medium bag radius is determined by the condition $\partial m_h^{*}/\partial R_h\vert_{R_h = R_h^{*}} = 0$. $z_h$ is related with the bag-model quantity that determined by the hadron mass in vacuum and the bag pressure $B = {\rm (170\, MeV)}^4$ fixed using the standard input of the QMC model for the nucleon in vacuum, $R_N = 0.8$ fm and $m_q = 5$ MeV. For the quarks inside the bag of the hadron $h$, the lowest positive eigenfunctions of the bag satisfy the boundary condition at the bag surface, $j_0 (x_{q}) =  \beta_{q}\, j_1 (x_{q})$, where $\beta_{q} = \sqrt{\Omega^*_{q} -(m^*_{q} R^*_h)/\Omega^*_{q} + (m^*_{q} R^*_h)}$ with $j_0$ and $j_1$ being the spherical Bessel functions. The scalar $\sigma$ and vector $\omega$ meson mean fields at the nucleon level can be related as
\begin{eqnarray}
  \label{eq:pionmed11}
  \omega = \frac{g_\omega\rho_B }{m_\omega^2},\,\,\,
  \label{eq:sigma}
  \sigma = \frac{4 g_\sigma C_N (\sigma)}{(2\pi)^3m_\sigma^2} \int d \bm{k}
  \frac{M_N^{*}(\sigma)\,\theta (k_F- | \bm{k}| ) }{\sqrt{M_N^{*2}(\sigma) + \bm{k}^2}}, 
\end{eqnarray}
where $C_N (\sigma)$ is defined as
\begin{eqnarray}
  C_N (\sigma) = \frac{-1}{g_\sigma (\sigma =0)} \left[ \frac{\partial M_N^{*} (\sigma )}{\partial 
      \sigma } \right].
\end{eqnarray}
Note that the value of $C_N (\sigma)$ is unity for the point-like nucleon. Both $C_N (\sigma)$ and $g_\sigma (\sigma)$ originate from the novel saturation properties of the QMC model and contain the dynamics of quark structure of the nucleon. By solving the scalar $\sigma$ mean field in Eq.~(\ref{eq:sigma}) self-consistently, the energy density is given by
\begin{eqnarray}
  \label{eq:pionmed12}
  \frac{E^{\rm tot}}{A} =\frac{4}{(2\pi)^3 \rho_B} \int d\bm{k} \, \theta (k_F - | \bm{k} |) 
  \sqrt{M_N^{*2} (\sigma) + \bm{k}^2}
  + \frac{m_\sigma^2 \sigma^2}{2\rho_B} + \frac{g_\omega^2 \rho_B}{2 m_\omega^2}.
\end{eqnarray}
All the coupling constants in Eq.~(\ref{eq:pionmed12}) are determined by fitting the binding energy of $15.7~\textrm{MeV}$ at the saturation density $\rho_0 =$ 0.16 fm$^{-3}$ for SNM. We then obtain that $g_\sigma^2 /4\pi =$ 5.393, $g_\omega^2/4\pi =$ 5.304, $B^{1/4} =$ 170, $M_N^* =$ 754.6, $K=$ 279.3, and $z_N =$ 3.295.
\begin{figure}[t]
  \centering
  \includegraphics[width=0.45\textwidth]{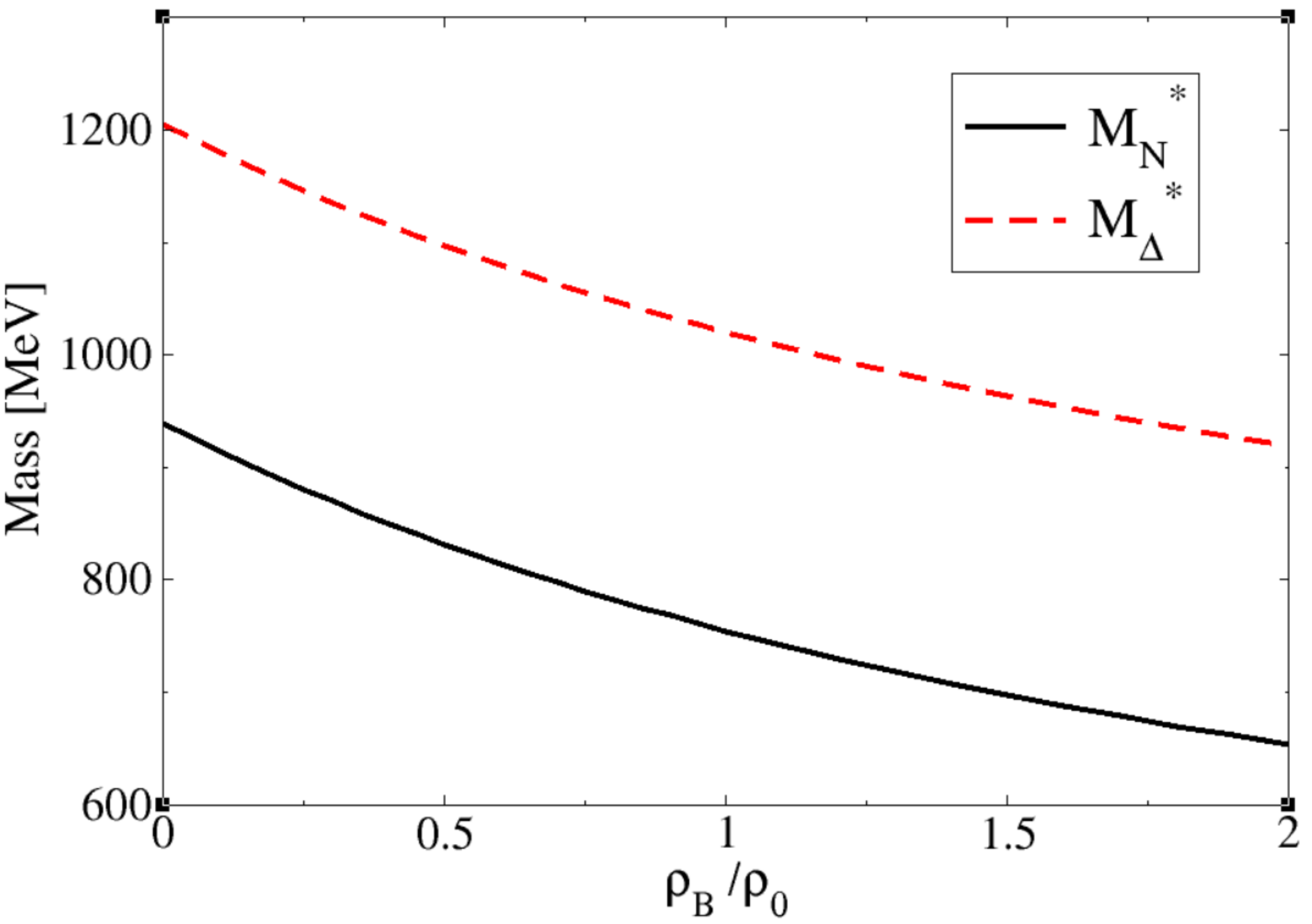}
  \caption{ \label{fig1b}(Color online) Effective baryon masses for the nucleon (solid) and $\Delta$ resonance (dashed) as functions of the baryon density $\rho_B / \rho_0$.}
\end{figure}

Using the obtained coupling constants, we calculate the effective baryon masses for $N$ and $\Delta$ resonance as shown in Fig.~\ref{fig1b}. The numerical results show that the effective nucleon mass decreases as the baryon density increases as expected. This behavior is followed by the effective delta mass which is consistent with other model calculations~\cite{Raduta:2021xiz,Dexheimer:2021sxs,Schurhoff:2010ph,Sen:2018tms,Sahoo:2018xeu}. These effective baryon masses will be used as inputs to the $\pi^{+}p$ cross section in medium. As for the pion, we do not consider its medium modifications, since we assumed that the incident and scattered pions are far outside the medium-effective region, such as the finite-size nuclei target for instance, as mentioned previously. 

\section{Numerical result}
\subsection{Elastic $\pi^{+}p$ scattering at $\rho_B=0$}
Here, we present our numerical results for the various scattering observables for the elastic $\pi^{+}p$ process in vacuum $(\rho_B=0)$. In our numerical calculations, we employ $f_{\pi NN} = 0.989$ obtained from the Nijmegen potential~\cite{Gasparyan:2003fp} and $f_{\pi N \Delta} = 2.127$ given by the experiment~\cite{Janssen:1996kx}. In principle, the full decay width of the $\Delta$ resonance, i.e., $\Gamma_\Delta$ depends on energies. In Ref.~\cite{Larionov:2003av}, the decay width is parameterized  as follows:
\begin{eqnarray}
  \label{eqN11}
  \Gamma_\Delta \left(\sqrt{s},M_{\pi,N,\Delta}\right)=\Gamma^0_\Delta \Big[ \frac{q \left(M_N,M_\pi,\sqrt{s}\right)}{q \left(M_N,M_\pi,M_\Delta \right)}\Big]^3 \frac{M_\Delta}{\sqrt{s}}\frac{\beta^2_0+q^2 \left(M_N,M_\pi,M_\Delta \right)}{\beta^2_0+q^2 \left(M_N,M_\pi,\sqrt{s}\right)},
\end{eqnarray}
where $\sqrt{s}$ and $\Gamma_{\Delta}^0$ are respectively the center-of-mass (cm) energy and Breit-Wigner width, respectively. The $\beta_0$ stands for the cut-off parameter, whereas $q$ denotes the three momentum of the intermediate particle, i.e., the $\Delta$ resonance:
\begin{eqnarray}
  \label{eqN13}
  q \left(M_b,M_c,M_a\right) &=& \frac{\sqrt{\left(M_a^2 + M^2_b - M^2_c\right)^2-4M_a^2 M_b^2}}{2M_a}.
\end{eqnarray}

First, using the decay width $\Gamma_\Delta$ in Eq.~(\ref{eqN11}), we calculate total cross section (TCS) as a function of $\textbf{p}_{\textrm{Lab}}$ and the angular-dependent differential cross section (DCS$_\theta$) for the elastic $\pi^{+}p$ scattering as a function of $\cos \theta$ in vacuum. Here, we choose $M_\Delta = 1215.5$ MeV and $\Gamma^0_\Delta = 94.0$ MeV, which are slightly different from their PDG values, to reproduce the experimental data.
\begin{figure}[t] 
  \centering
  \stackinset{r}{6cm}{t}{0.5cm}{(a)}{\includegraphics[width=0.45\textwidth]{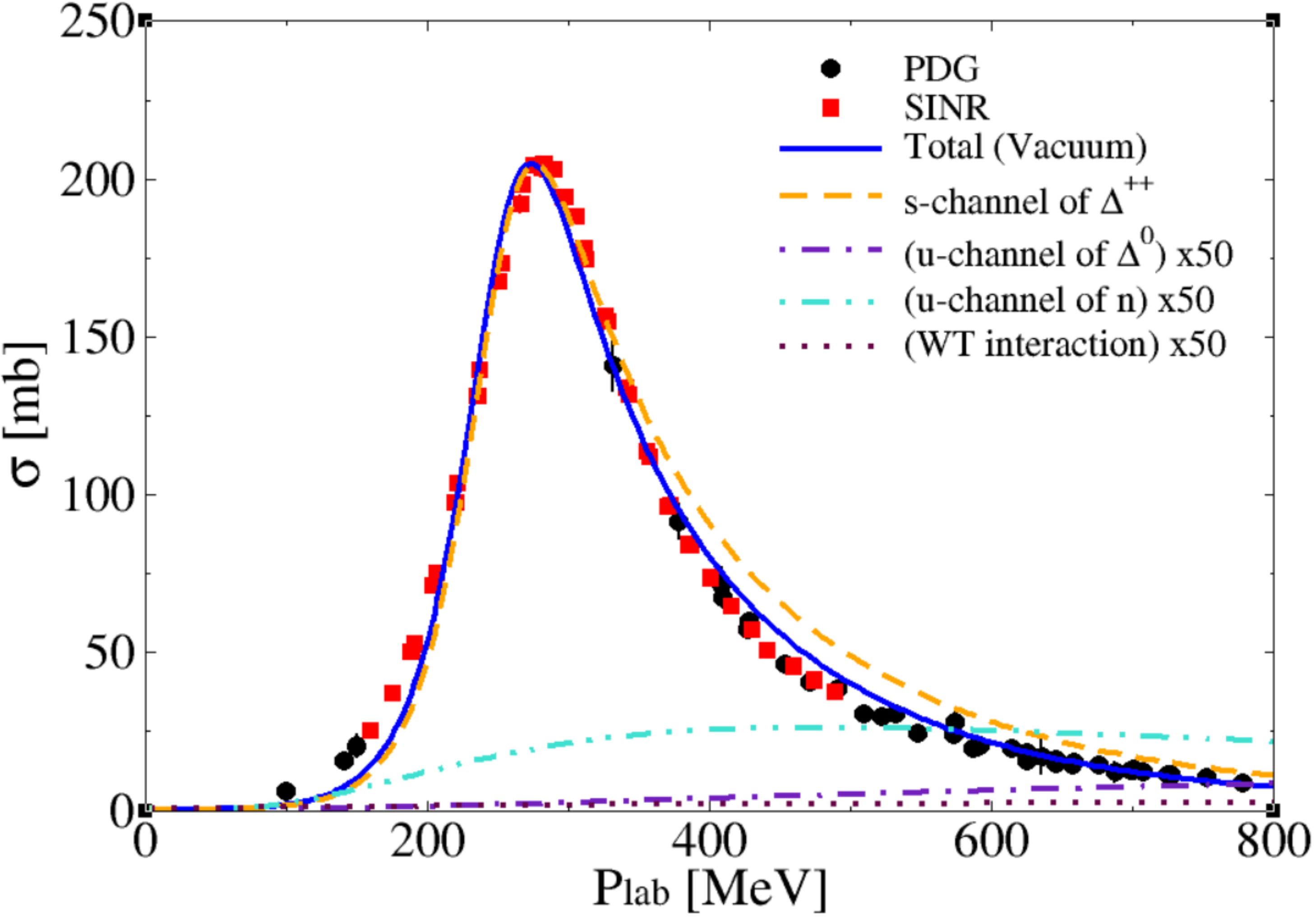}}
  \stackinset{r}{1cm}{t}{0.5cm}{(b)}{\includegraphics[width=0.45\textwidth]{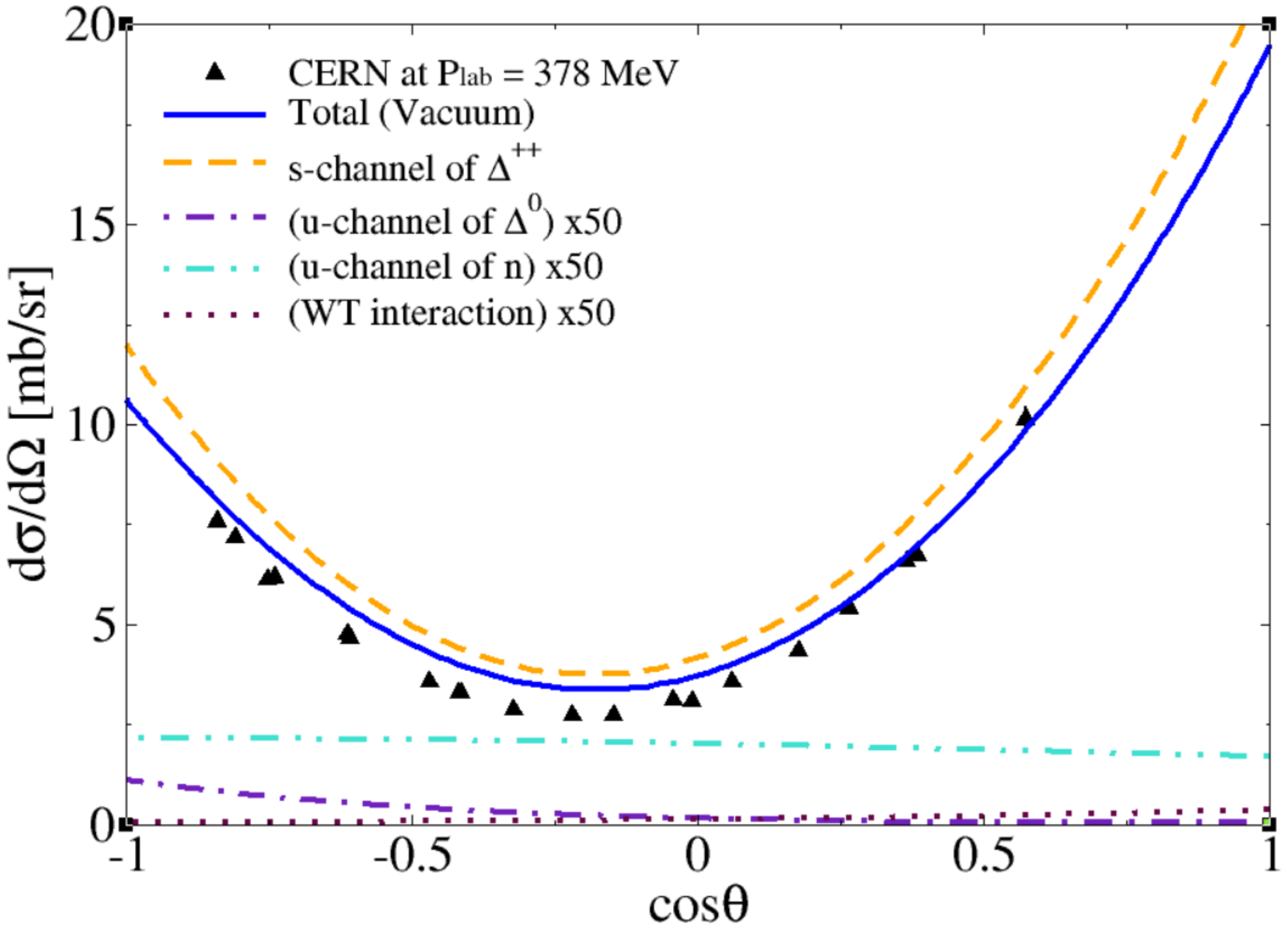}}
  \caption{ \label{fig1}(Color online) Total cross section (TCS) of the elastic $\pi^{+}p$ scattering as a function of $\textbf{p}_\mathrm{lab}$ (a) and angular-dependent differential cross section (DCS$_{\theta}$) as a function of $\cos\theta$ (b) for $\rho_B=0$ (vacuum). Here, we employ $M_\Delta =$ 1215.5 MeV and  $\Gamma^0_\Delta =$ 94.0 MeV. The experimental data for TCS are taken from the Particle Data Group (PDG) compilation~\cite{PDG2020} (circle) and Swiss Institute of Nuclear Research (SINR)~\cite{Pedroni:1978it} (square). Those for DCS$_{\theta}$ from Ref.~\cite{Bussey:1973gz} (triangle). Each contribution is separately shown as well.}
\end{figure}

The panel (a) of Fig.~\ref{fig1} shows the comparison between TCS (solid line) and the experimental data, which are taken from the Particle data group (PDG) compilation~\cite{PDG2020} (circle) and Swiss Institute of Nuclear Research (SINR)~\cite{Pedroni:1978it} (square). It turns out that the numerical results for TCS is in a good agreement with the experimental data. Also each channel contribution for TCS is given their separately. It can be clearly seen that the dominant contribution of $\Delta$(1232) in the $s$ channel is crucial to produce the peak in the cross section in the vicinity of the threshold.

The panel (b) of Fig.~\ref{fig1} shows the numerical results for the differential cross section as a function of $\cos\theta$ (DCS$_{\theta}$), in which $\theta$ denotes the cm-frame angle for the scattered pion. Our result fits relatively well with the existing data from CERN at $\textbf{p}_{\textrm{lab}} =$ 378 MeV (triangle)~\cite{Bussey:1973gz}, showing in a typical $d$-wave scattering, $\propto\textbf{k}'\cdot\textbf{k}$, due to the $\Delta$ resonance. In the backward-scattering region, however, the numerical result slightly overestimates the data. It is expected that other baryon resonances for the $u$-channel neglected in the present work for brevity are necessary to improve the result, and those improvements will be taken into account in the future works.

\subsection{Elastic $\pi^{+}p$ scattering at $\rho_B\ne0$}
\begin{figure}[t]
\includegraphics[width=0.45\textwidth]{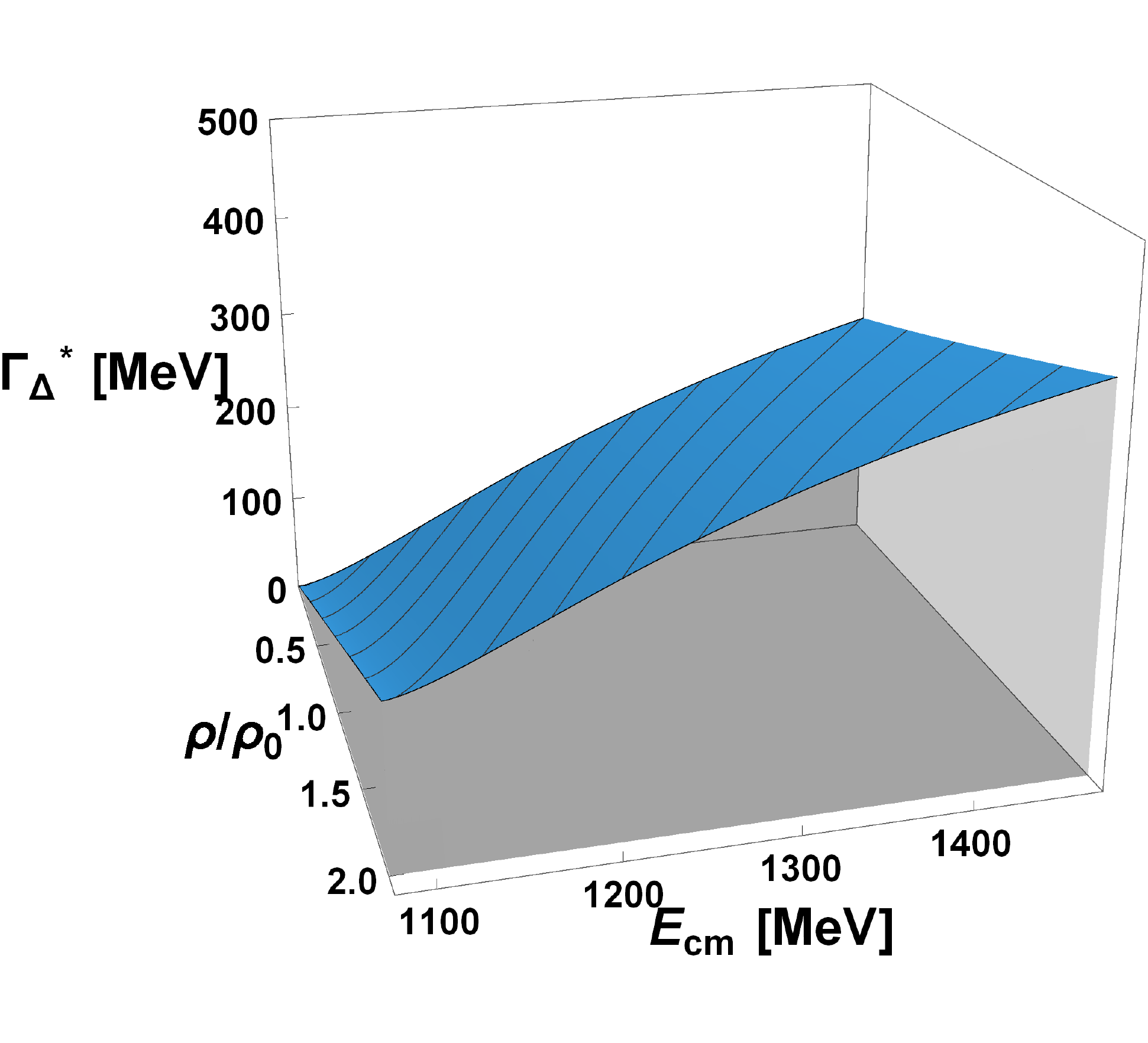}
\caption{(Color online) In-medium $\Delta$-resonance full decay width $\Gamma^*_\Delta$ as a function of $E_{\textrm{cm}}$ and $\rho_B/\rho_0$ from Eq.~(\ref{eqN12}).}
\label{fig2}
\end{figure}

Now we are in a position to consider the physical observable at finite baryon density. Note that, in addition to the hadron mass modifications as discussed in the previous Section, the decay width for the $\Delta$ resonance is modified as well in medium as follows:
\begin{eqnarray}
  \label{eqN12}
  \Gamma^*_\Delta \left(\sqrt{s^*} \right) =  \Gamma_\mathrm{sp} \left( \frac{\rho_B}{\rho_0} \right)
  +   \Gamma_\Delta \left(\sqrt{s^*},M^*_{\pi,N,\Delta}\right),
\end{eqnarray}
where $\rho_B$ and $\Gamma_\mathrm{sp}$ are the baryon density and spreading width of the resonance in medium, respectively. In this calculation, we take the values of the $\Gamma_\mathrm{sp}$ = 80 MeV, which determined by fitting to the medium quantity at finite density~\cite{Larionov:2003av,Hirata:1978wp,Oset:1987re}, and the $\beta_0 = 200$ MeV in Eq.~(\ref{eqN11}) is a cut-off parameter~\cite{Larionov:2003av}. The numerical result for the in-medium $\Delta$-resonance full decay width $\Gamma^*_\Delta$ in Eq.~(\ref{eqN12}) as a function of $E_{\textrm{cm}}$ and $\rho_B/\rho_0$ is depicted in Fig.~\ref{fig2}. It shows that $\Gamma^*_\Delta$ increases with respect to the energy as well as the density, indicating that the \textit{dissociating} $\Delta$ resonance in dense medium, due to the weaker interaction strength~\cite{Ghosh:2016hln,Cui:2020fhr}.

\begin{figure}[t]
  \centering
  \includegraphics[width=0.45\textwidth]{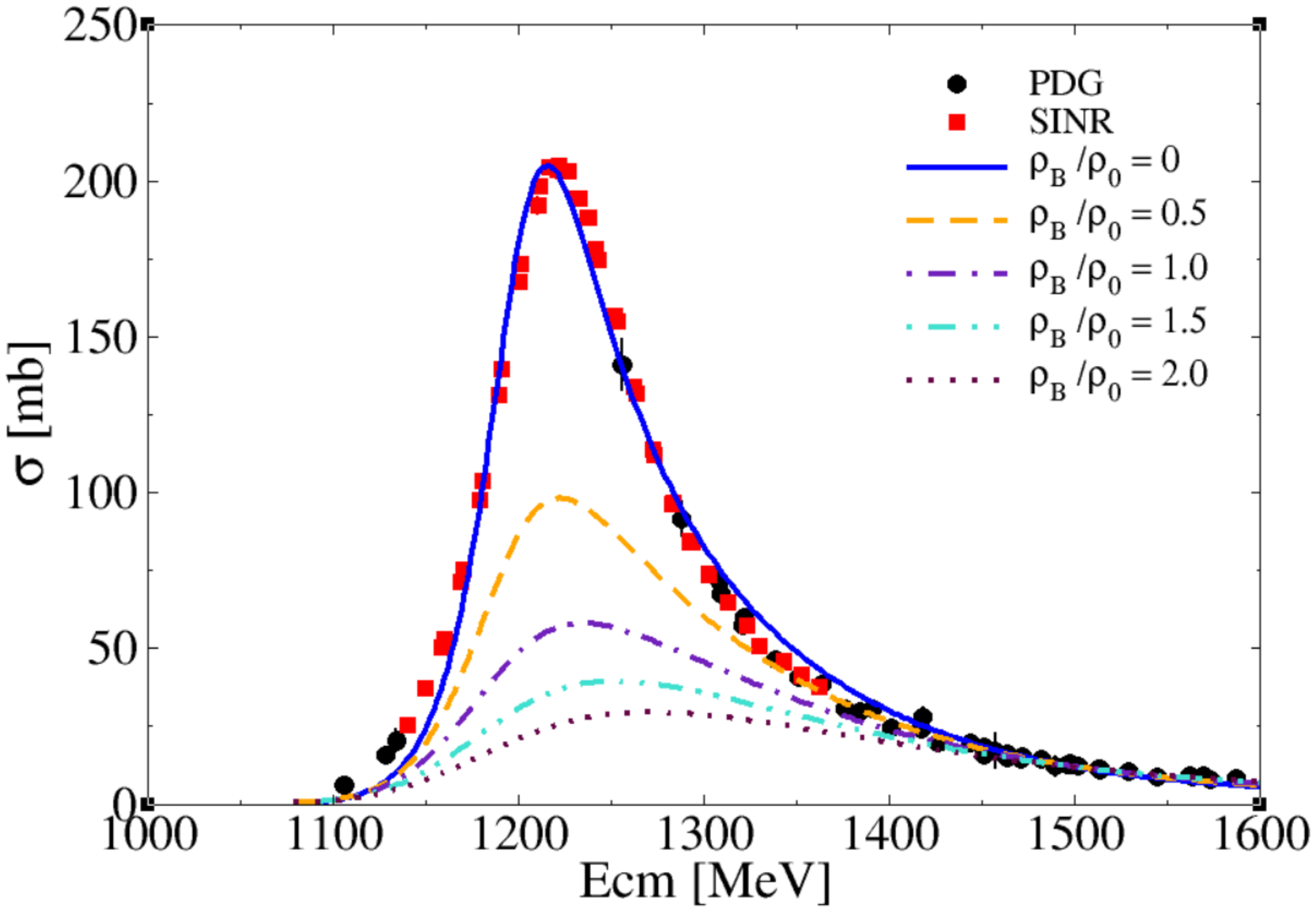}
  \caption{ \label{fig3}(Color online) Total cross section for the $\pi^{+}p$  elastic scattering as a function of $E_{\textrm{cm}}$ for various baryon densities. The circle and square denote the experimental data from Refs.~\cite{Pedroni:1978it,PDG2020} for vacuum.}
\end{figure}

Using the density-dependent width and mass, i.e., $\Gamma^*_\Delta$ and  $M^*_B$, we calculate TCS as a function of $E_{\textrm{cm}}$ for various baryon densities of $\rho_B/\rho_0=(0-2)$ in Fig.~\ref{fig3}. The experimental data for the vacuum case are taken again from Refs.~\cite{Pedroni:1978it,PDG2020}. As the density increase, the peak of the $\Delta$ resonance gets diminished as well as broadened, as expected from the numerical results shown in Fig.~\ref{fig2}. Physically, this observation can be understood by that the interaction strength between the pion and nucleon becomes weak in medium, resulting in the lower production rate of the $\Delta$ resonance. Moreover, it is worth noting that the peak position of the resonance smoothly moves to the higher cm energy. Similar tendencies are reported in other theoretical model calculations~\cite{Ghosh:2016hln,Cui:2020fhr}.

\begin{figure*}[t]
  \centering
  \stackinset{r}{4cm}{t}{0.5cm}{(a)}{\includegraphics[width=0.33\textwidth]{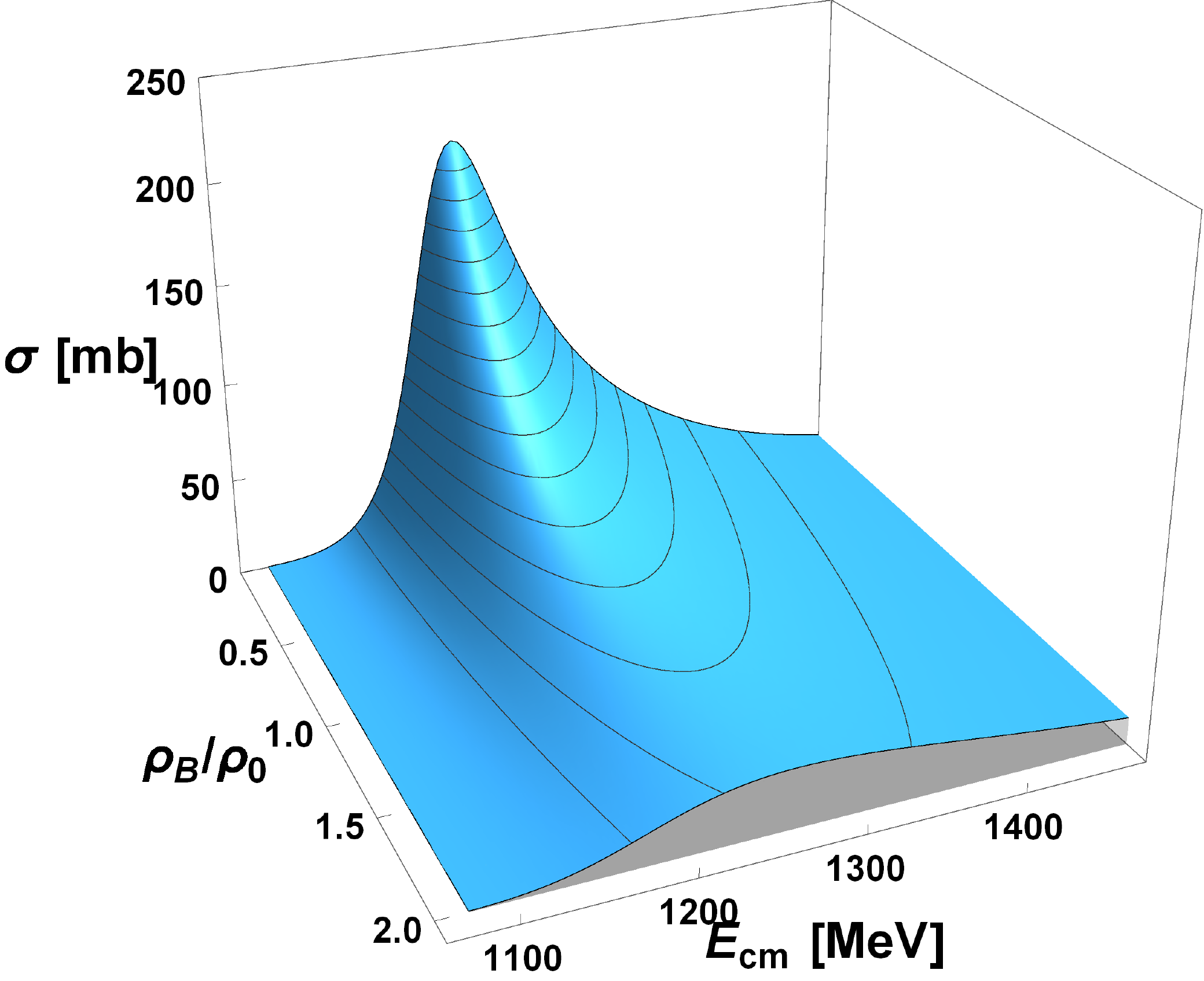}}
  \stackinset{r}{4cm}{t}{0.5cm}{(b)}{\includegraphics[width=0.33\textwidth]{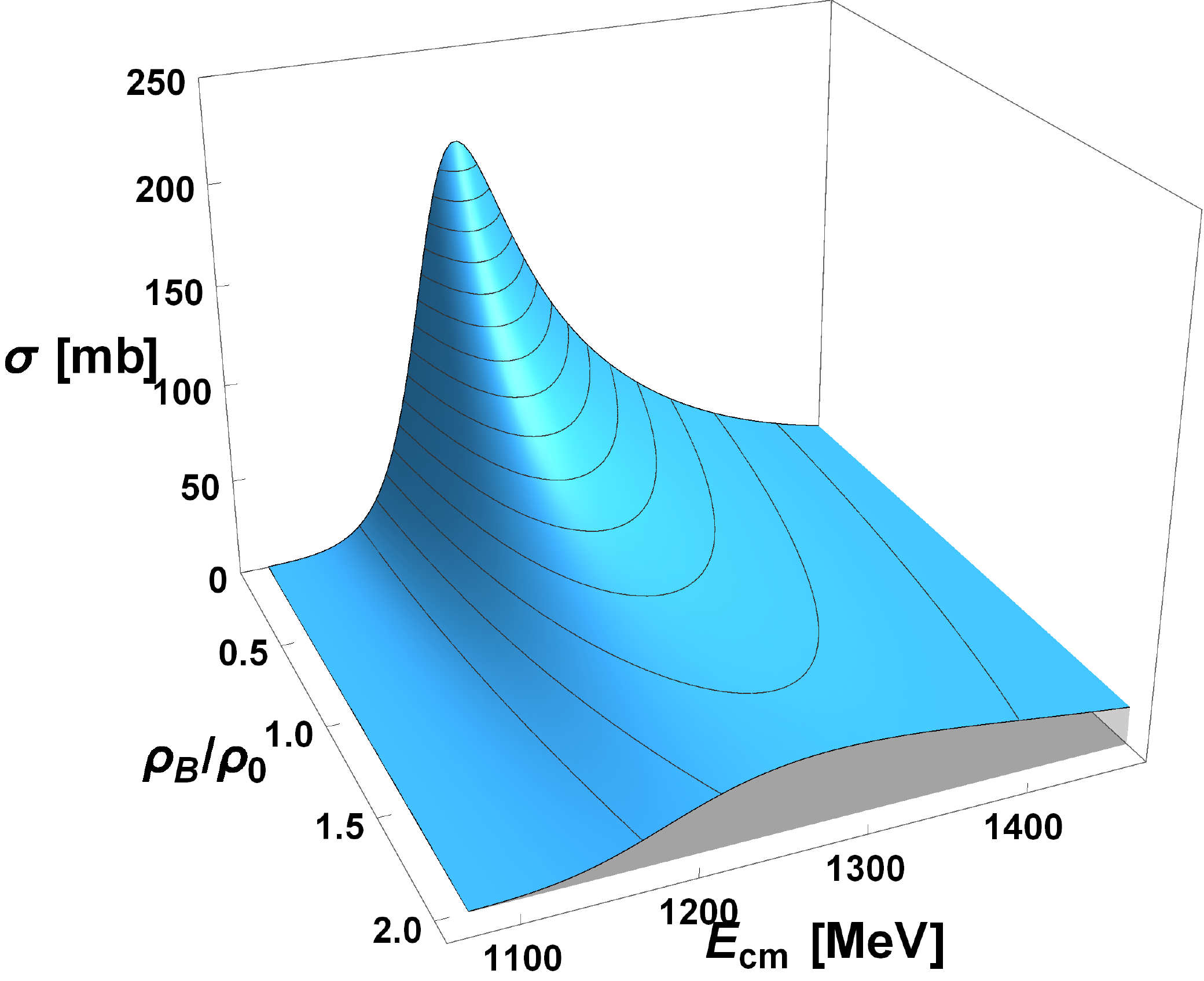}}
  \stackinset{r}{4cm}{t}{0.5cm}{(c)}{\includegraphics[width=0.33\textwidth]{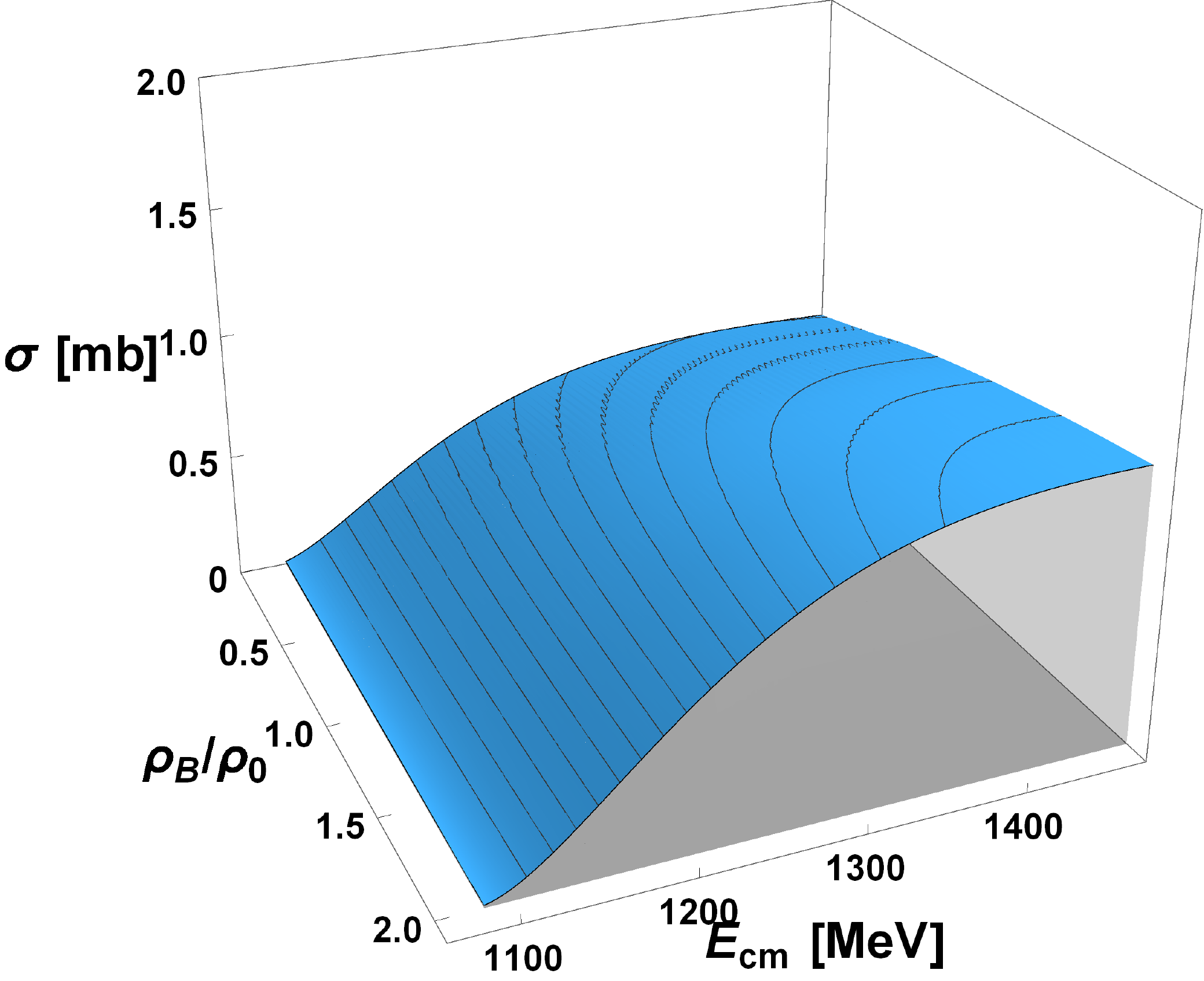}}
  \caption{ \label{fig4}(Color online) Total cross section for the elastic $\pi^+p$ scattering as a function of $E_{\textrm{cm}}$ and $\rho_B/\rho_0$ for the total (a), $s$-channel $\Delta$(1232) (b), and background (c) contributions.}
\end{figure*}

In Fig.~\ref{fig4}, we depict the total (a), $s$-channel $\Delta$(1232) (b), and background (c) contributions separately for TCS as a function of $E_{\textrm{cm}}$ and $\rho_B/\rho_0$. As shown there, the $\Delta$-resonance contribution dominates the scattering process and decreases rapidly with respect to the density. On the contrary, the BKG contribution slightly increases as a function of the density, but it is almost negligible. 

\begin{figure}[t]
  \centering
\stackinset{r}{2cm}{t}{0.5cm}{(a)}{  \includegraphics[width=0.45\textwidth]{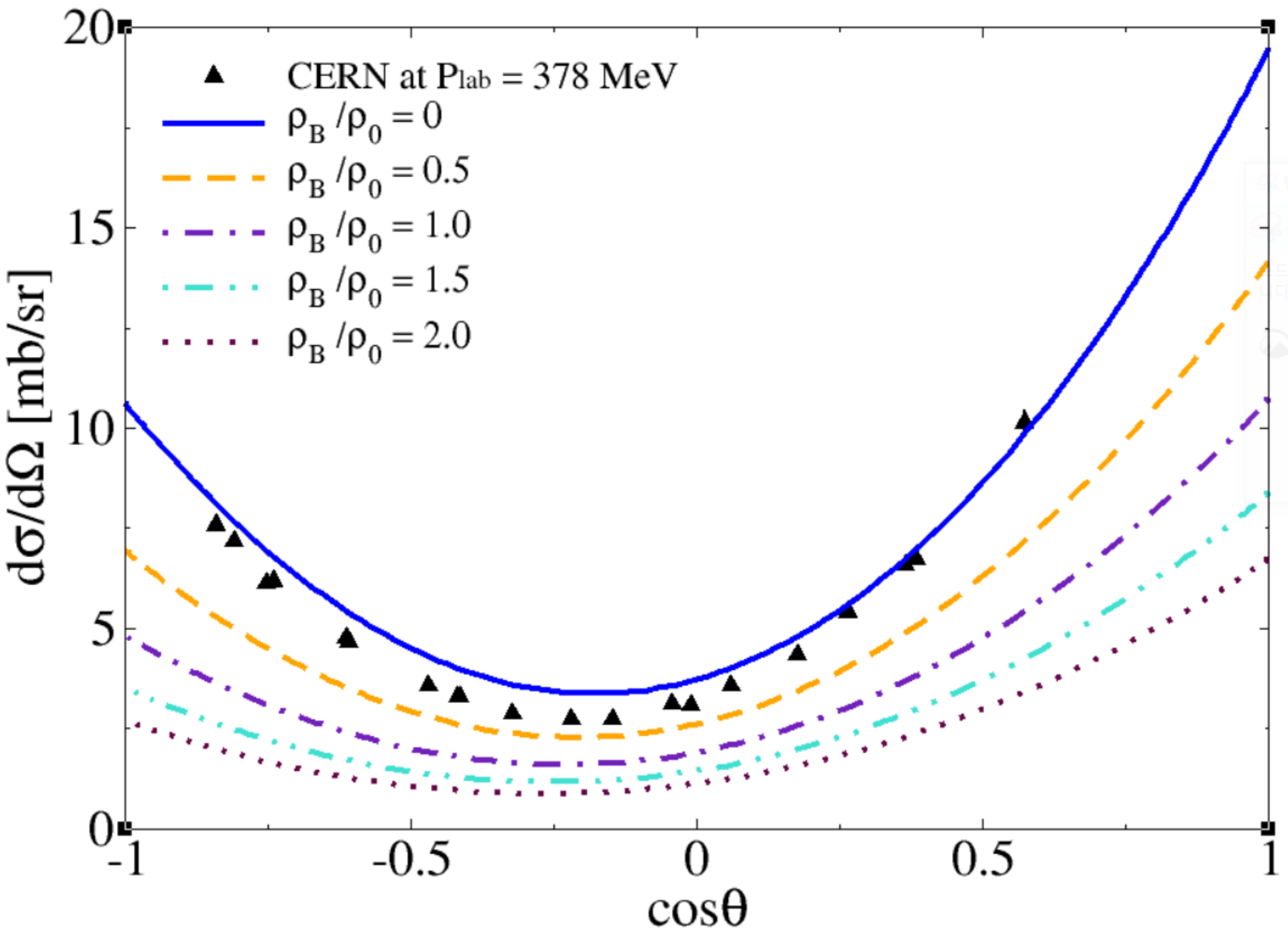}}
 \stackinset{r}{2cm}{t}{0.5cm}{(b)}{ \includegraphics[width=0.45\textwidth]{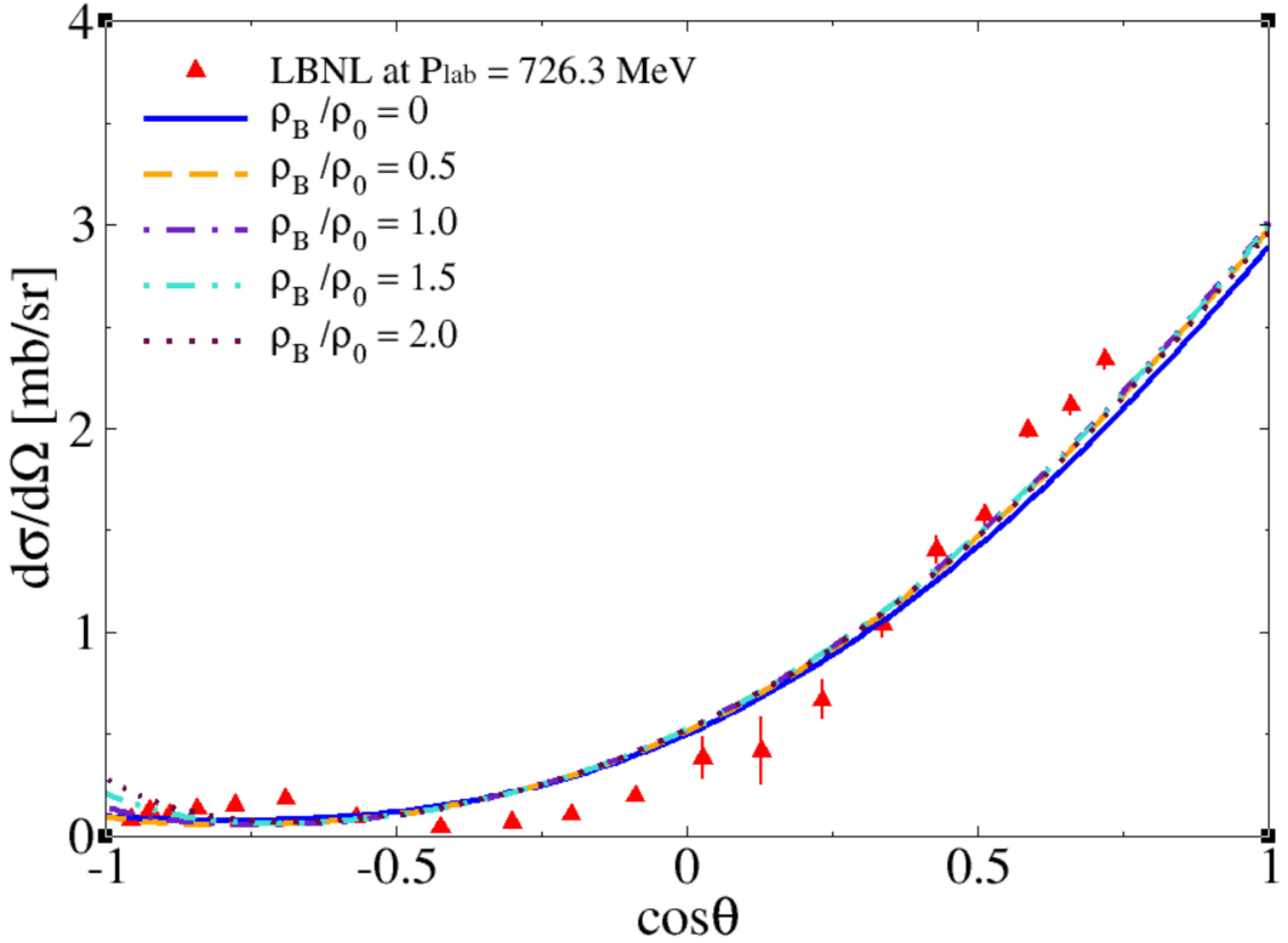}}
  \caption{ \label{fig5}(Color online) Differential cross section DCS$_{\theta}$ for the elastic $\pi^+p$ scattering as a function of $\cos\theta$ for $\rho_B/\rho_0=(0-2)$ at $\textbf{p}_{\textrm{lab}}$ = 378 MeV (a) and  $726.3$ MeV (b). The experimental data is taken from Ref.~\cite{Bussey:1973gz,Ogden:1965zz}.}
\end{figure}

The numerical results for the angular dependent DCS$_{\theta}$ are shown in the panel (a) and (b) of Fig.~\ref{fig5} for $\rho_B/\rho_0=(0-2)$ at $\textbf{p}_{\textrm{lab}} = 378$ MeV and $726.3$ MeV, respectively. The experimental data is taken from Ref.~\cite{Bussey:1973gz,Ogden:1965zz} for $\rho_B=0$. For $\textbf{p}_{\textrm{lab}} = 378$ MeV at which the $\Delta$ resonance dominates the process ($E_\mathrm{cm}\approx1290$ MeV), the angular dependence shows a typical $d$-wave curve, due to the dominant $\Delta$ resonance, showing the strong forward and mild backward scattering enhancements, and the strength gets diminished with respect to the density as expected from Fig.~\ref{fig4}. As the momentum increases up to $\textbf{p}_{\textrm{lab}} = 726.3$ MeV ($E_\mathrm{cm}\approx1510$ MeV), where the $\Delta$-resonance contribution becomes compatible with the BKGs, and interferes with the $u$-channel BKG contributions destructively, resulting in the forward scattering enhancements as shown in the panel (b) of Fig.~\ref{fig5}.

Interestingly, at $\textbf{p}_{\textrm{lab}} = 726.3$ MeV, the curves remain almost the same even with the density changes for $\rho_B/\rho_0=(0-2)$, since the density dependence becomes weak and less visible in the relatively higher energy regions as shown in the TCS results as indicated in Figs.~\ref{fig3} and \ref{fig4}. Note that the experimental data for $\textbf{p}_{\textrm{lab}}$ = 726.3 MeV~\cite{Ogden:1965zz} is well reproduced by the theory, indicating the validity of the present model calculation. We verified that the backward scattering is enhanced as the momentum goes higher than $800$ MeV, since the $\Delta$-resonance contribution almost disappears and the $u$-channel BKG contributions start to be important. 

\begin{figure}[t]
\stackinset{r}{5cm}{t}{0.5cm}{(a)}{\includegraphics[width=0.45\textwidth]{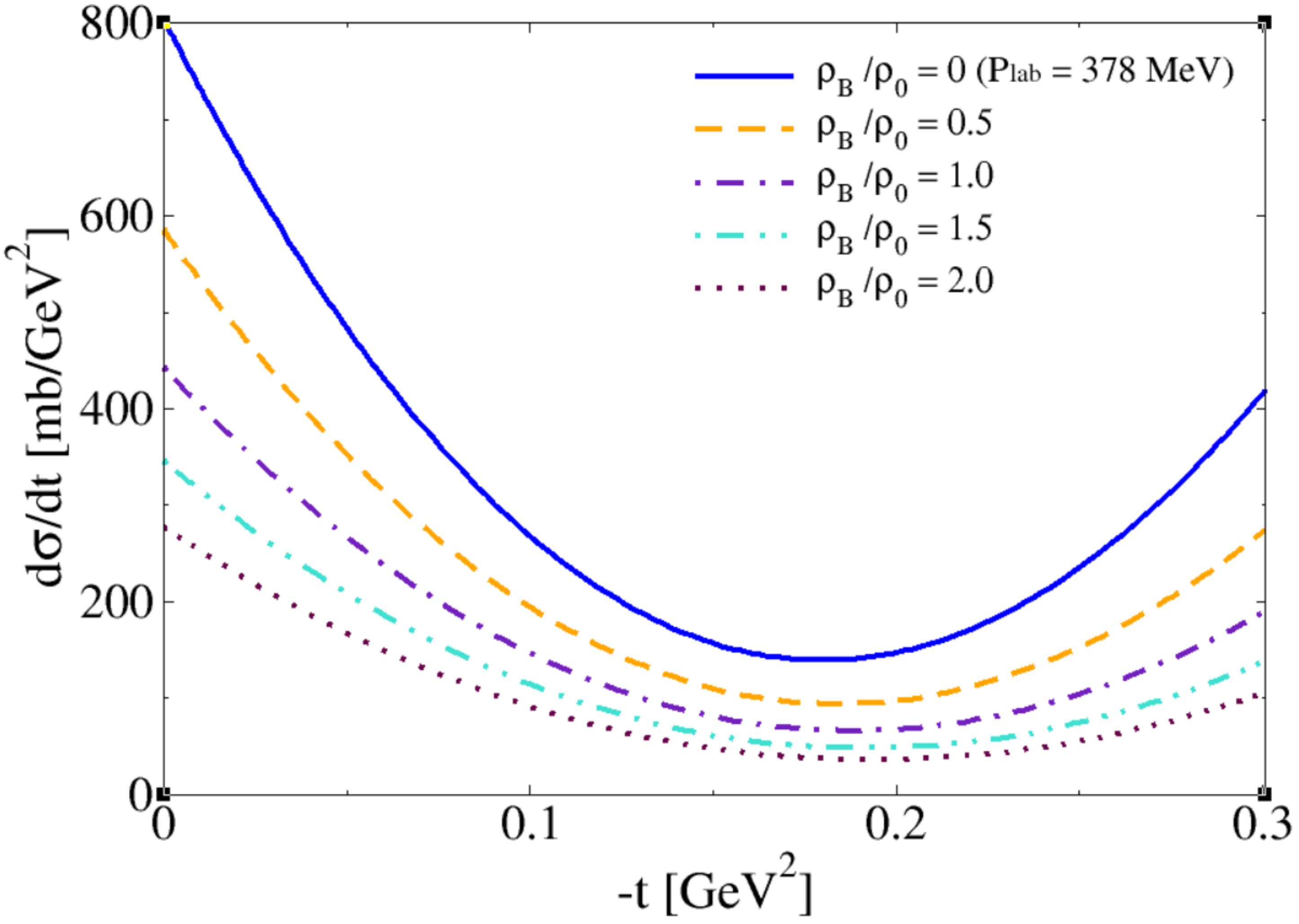}}
\stackinset{r}{5cm}{t}{0.5cm}{(b)}{\includegraphics[width=0.45\textwidth]{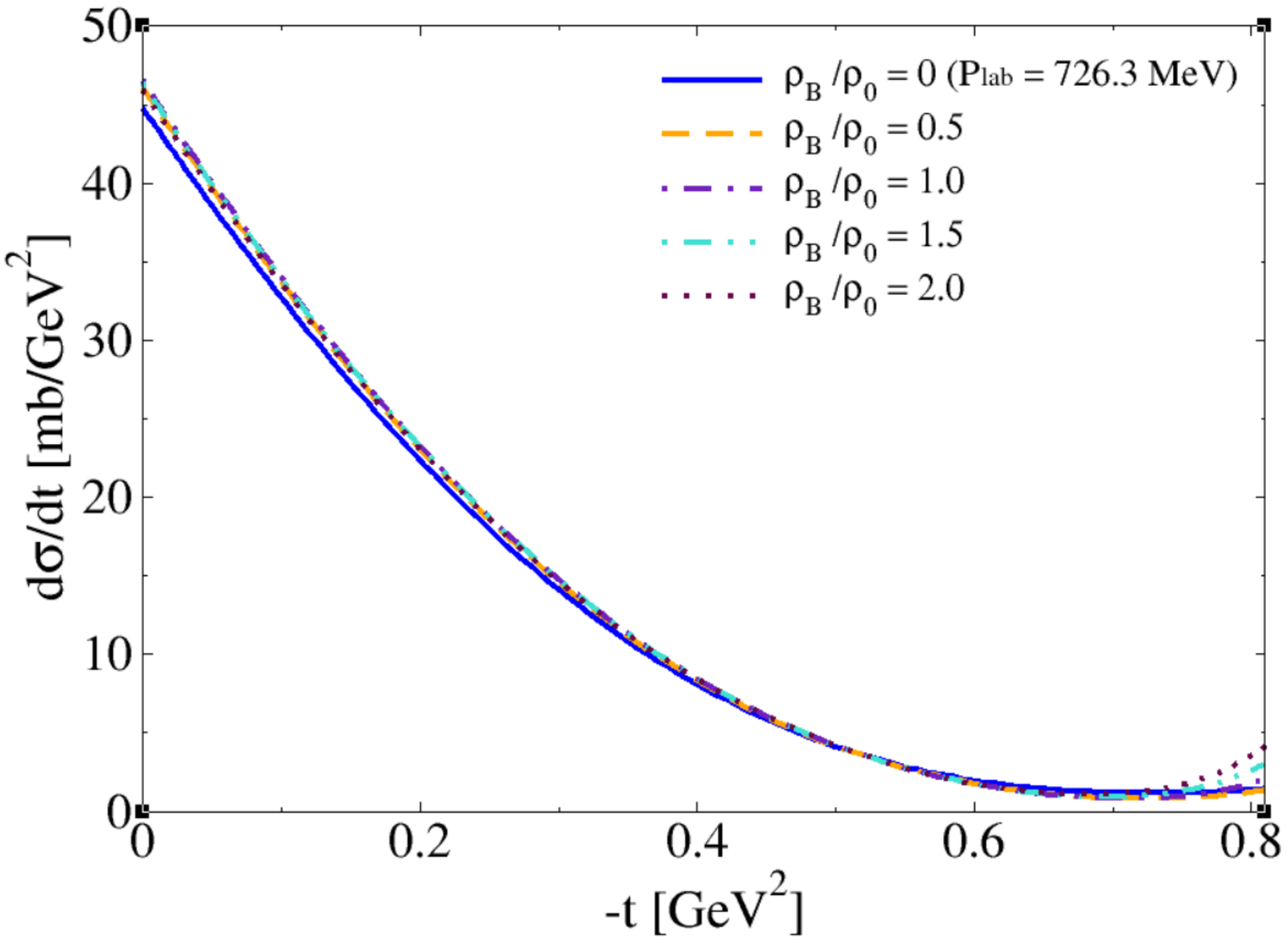}}
\caption{(Color online) $t$-dependent differential cross section (DCS$_{t}$) for the elastic $\pi^{+}p$ scattering as a function of $-t$ for $\rho_B/\rho_0=(0-2)$ at $\textbf{p}_{\textrm{lab}}$ = 378 MeV (a) and  $726.3$ MeV (b).}
\label{fig7}
\end{figure}

The numerical results for the $t$-dependent differential cross section (DCS$_{t}$) as a function of $-t$ for the densities at $\textbf{p}_{\textrm{lab}}$ = 378 MeV and  $726.3$ MeV are given in the panel (a) and (b) of Fig.~\ref{fig7}, respectively. At $\textbf{p}_{\textrm{lab}}$ = 378 MeV, we observe the strong forward- and backward-scattering enhancements as already shown in DCS$_{t}$, and similar tendencies are found for the higher momentum. The density dependence can be understood in the same way.  

\begin{figure}[t]
  \centering
 \stackinset{r}{0.5cm}{t}{1.3cm}{(a)}{ \includegraphics[width=0.45\textwidth]{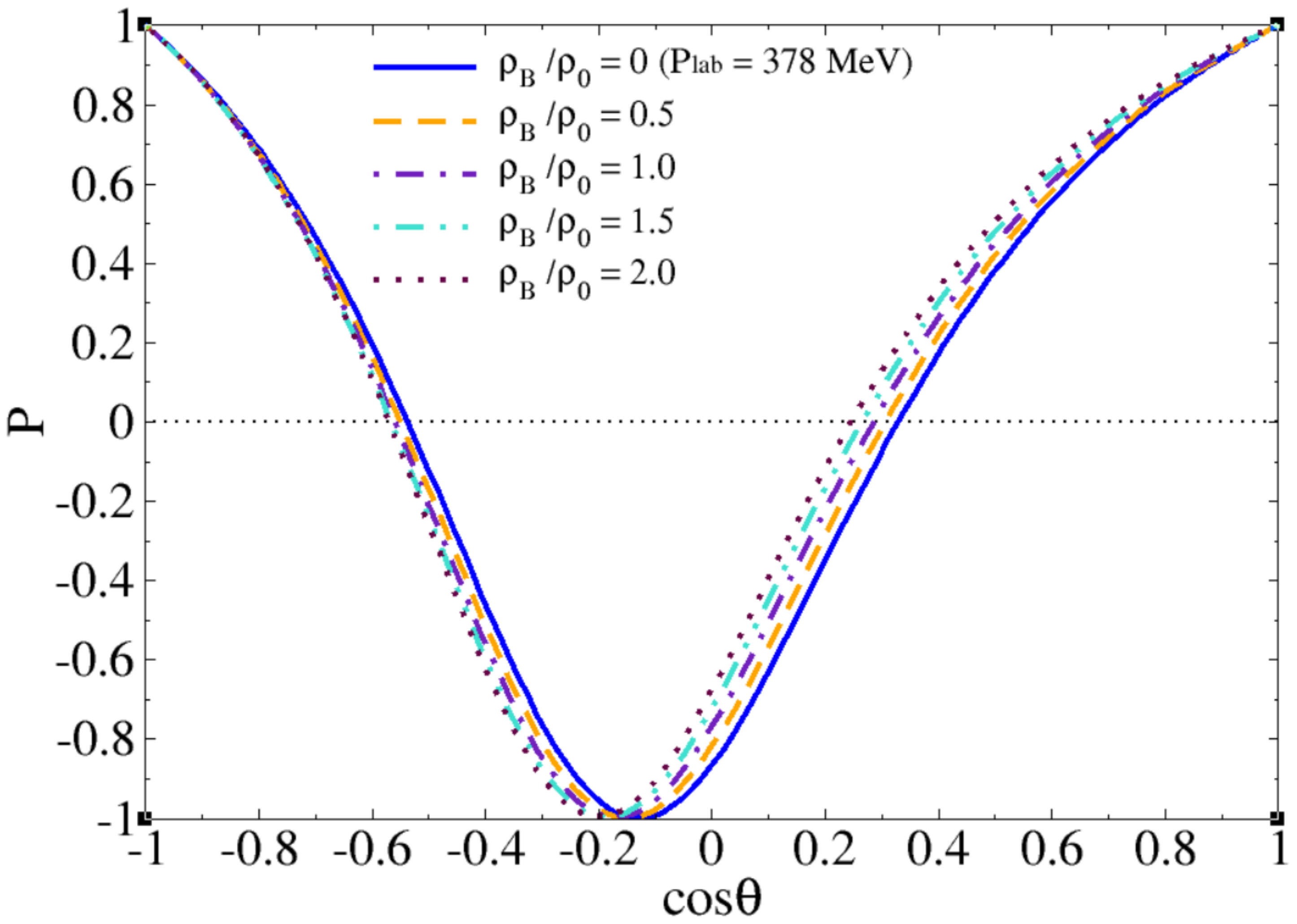}}
 \stackinset{r}{0.5cm}{t}{1.3cm}{(b)}{ \includegraphics[width=0.45\textwidth]{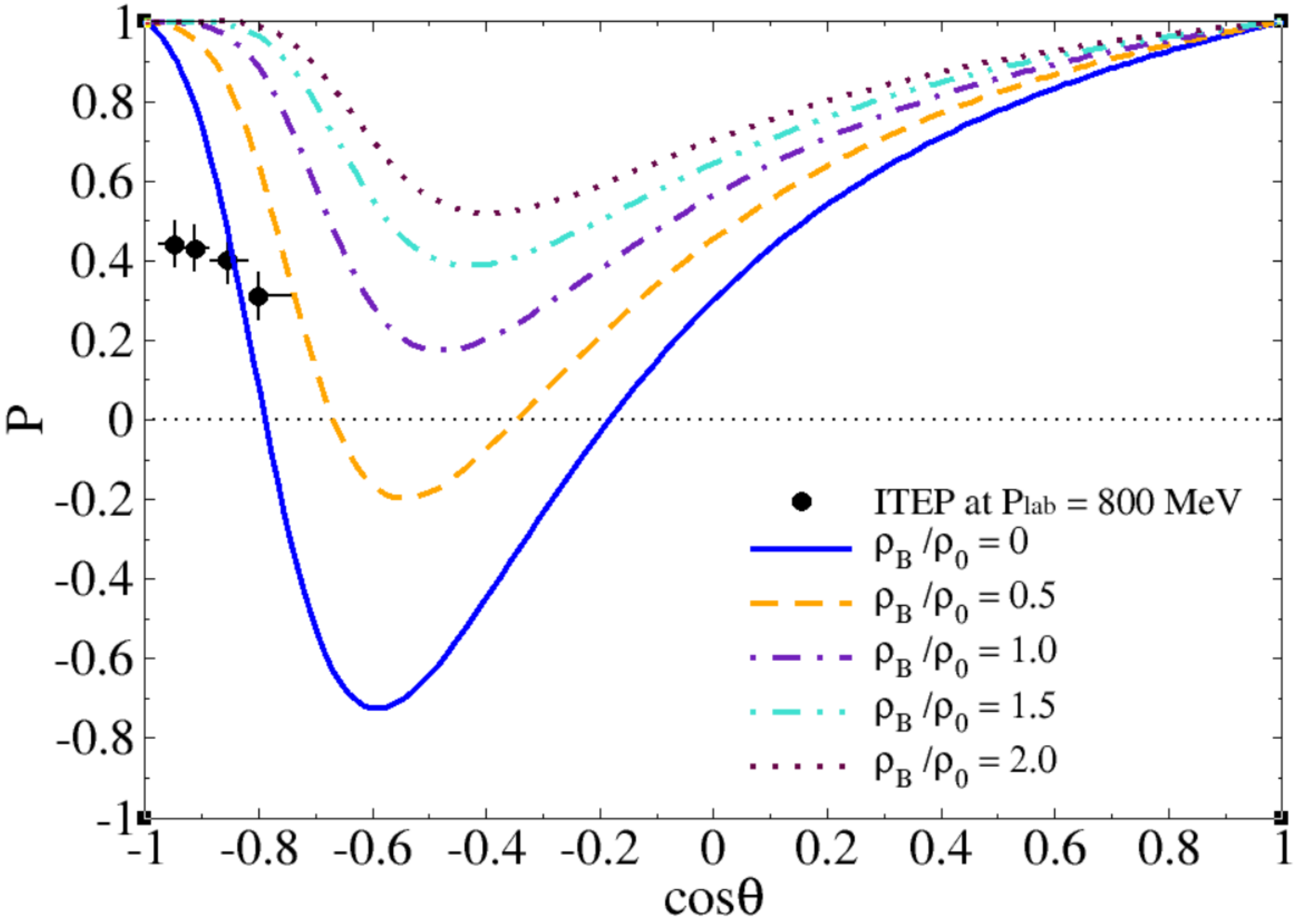}}
  \caption{ \label{fig6}(Color online) Target- and recoil-proton spin asymmetry $P$ in Eq.~(\ref{eqN14}) for the elastic $\pi^{+}p$ scattering as a function of $\cos \theta$ for $\rho_B/\rho_0=(0-2)$ at $\textbf{p}_{\textrm{lab}}$ = 378 MeV (a) and  $800$ MeV (b). The experimental data from the ITEP-PNPI collaboration at $\textbf{p}_\mathrm{lab}=800$ MeV~\cite{ITEP-PNPI:2008cmv}.}
\end{figure}

Finally, we compute the target- and recoil-proton spin asymmetry $P$ defined as follows:
\begin{eqnarray}
  \label{eqN14}
P&= & \frac{\sum_\mathrm{spin}\left[\left( \frac{\partial \sigma}{\partial \Omega}\right)_\mathrm{S}-\left(\frac{\partial\sigma}{\partial\Omega}\right)_\mathrm{O}\right]}{\sum_\mathrm{spin}\left[\left( \frac{\partial \sigma}{\partial \Omega}\right)_\mathrm{S}+\left(\frac{\partial\sigma}{\partial\Omega}\right)_\mathrm{O}\right]},
\end{eqnarray}
where the subscripts S and O stand for that the target- and recoil-proton spins are aligned in the same and opposite quantization axes, respectively. This physical quantity represents the short-range spin correlation inside the medium, i.e., a nuclei target for instance. In the panel (a)  and (b) of Fig.~\ref{fig6}, we show the numerical results of $P$ as a function of $\cos\theta$ for the densities at $\textbf{p}_{\textrm{lab}}$ = 378 MeV and $800$ MeV, respectively. The curve shape can be easily understood by the spin statistics. When the spins of the target and recoil proton are in the same direction, the intermediate $\Delta$ resonance is in its $S=1/2$ spin state and its angular dependence $(d\sigma/d\Omega)_\mathrm{S}$ is given by $\sim(\mathrm{const.}+\cos^2\theta)$. As for the opposite direction, $(d\sigma/d\Omega)_\mathrm{O}$ is described by $\sim\sin^2\theta$. Hence, the combination of these two angular dependencies results in that curve shape commonly for the two momenta. 

Interestingly, however, the density dependence is quite different for the momenta as seen in Fig.~\ref{fig6}. The reason for this finding can be explained as follows: At the lower momentum, the scattering process is dominated by the $\Delta$-resonance contribution. Then, as the density increases, $(d\sigma/d\Omega)_\mathrm{O}$ and $(d\sigma/d\Omega)_\mathrm{S}$ decrease in the same rate, resulting in the stable $P$ with respect to $\rho_B$ as shown in the panel (a) of Fig.~\ref{fig6}. On the contrary, as for the higher momentum, although the $\Delta$-resonance contribution is still larger than other ones, but it starts to compete with the BKG contributions, resulting in the obvious density dependence as shown in the panel (b) of the Fig.~\ref{fig6}. It is worth mentioning that the experimental data from the ITEP-PNPI collaboration for $P$ at $\textbf{p}_\mathrm{lab}=800$ MeV~\cite{ITEP-PNPI:2008cmv}  is qualitatively well reproduced by the theory for $\rho_B=0$. 

\section{Summary and future perspectives}
In the present work, we investigated the elastic $\pi N$ scattering process, which is dominated by the $\Delta(1232)$-resonance contribution, for the $I=3/2$ channel in dense medium ($\rho_B\ne0$) in order to examine the in-medium modification of the hadron scattering process. The modification is caused by the hadron-property (mass and width for instance) changes in medium, due to the partial restoration of the spontaneous breakdown of chiral symmetry (SBCS). For this purpose, we employed the effective Lagrangian approach at the tree-level Born approximation to compute the scattering amplitudes for the relevant Feynman diagrams. To simplify the background contributions, we made use of the Weinberg-Tomozawa contact interaction, which mimics the vector meson exchange in the $t$ channel in the low-energy region, in addition to the nucleon- and $\Delta$-intermediate state diagrams in the $u$ channel. The phenomenological form factors were taken into account accommodating the spatial extension of the hadrons. The quark-meson coupling (QMC) model and linear-density approximation are employed to describe the medium modifications for the in-medium baryon properties, such as the baryon masses ($M^*_{N,\Delta}$) and $\Delta$ full decay width ($\Gamma^*_\Delta$). Below, we list important observations in the present work:
\begin{enumerate}
\item Total and differential cross sections are qualitatively well reproduced in comparison with the experimental data. Again, we confirmed that the scattering process is dominated by the $\Delta$ resonance in the $s$ channel. The angular dependence shows the strong forward and mild backward scattering enhancements in the vicinity of the $\Delta$ mass region, due to the $\Delta$-dominated $d$-wave scattering. As the energy increases, the $\Delta$-resonance and $u$-channel BKG contributions compete destructively each other, resulting more enhancements in the forward scattering.  
\item As for the symmetric nuclear matter (SNM), the nucleon and $\Delta$-resonance masses are decreasing with the same rate by the $\sigma$-meson exchange interaction with respect to the baryon density $\rho_B$ from the QMC model. The full decay width of $\Delta$ resonance, $\Gamma^*_\Delta$ gets wider as the density and energy $E_\mathrm{cm}$ increase from the linear-density approximation. This observation indicates that the $\pi N$ interaction becomes weak in medium, i.e., hard to produce the $\Delta$ resonance.  
\item The $\Delta$-resonance invariant-mass spectrum shown in the total cross section becomes dissolved with respect to the density, according to the weak $\pi N$ interaction in medium, and the peak position slightly moves to the higher energy regions. As for the angular-dependent differential cross section, a similar density dependence is shown in the vicinity of the $\Delta$-resonance region, while the density dependence almost disappears beyond the region, since the BKG contributions compete and overcome the $\Delta$-resonance one. We observed that the $t$-dependent differential cross section also depicts a similar tendency to the angular-dependent one as well. 
\item Finally, we compute the target-recoil proton-spin asymmetry, which signals the short-range spin correlation in medium. We find that the asymmetry follows a simple spin statistics, and the resulting curve shapes can be understood by the combination of the $S=1/2$ and $S=3/2$ spin states of the intermediate $\Delta$ resonance. In the vicinity of the $\Delta$-resonance mass region, the asymmetry is dominated by it and insensitive to the density changes. On the contrary, as for the regions beyond $\Delta$-resonance mass, the density dependence becomes finite and obvious, since the $\Delta$-resonance and BKG contributions start to compete. 
\end{enumerate}

We expect that these observations will shed light on the understandings for the $\Delta$-resonance contribution in the relativistic heavy-ion collision as well as the neutron star EoS.  As mentioned, however, the present setup for the scattering in medium is quite ideal and different from the reality. An actual nuclei ($A$) target has a distance-dependent density profile $\rho_A(r)$, not the uniform density, being given by the Woods-Saxon type distribution for instance. Hence, the elastic $\pi A$ scattering cross section can be obtained by $d\sigma^\mathrm{elas.}_{\pi A}(\sqrt{s})\sim\,\sigma^\mathrm{elas.}_{\pi N}(\sqrt{s},\rho_A) \times \rho_A dV$, using the present results for $\sigma^\mathrm{elas.}_{\pi N}(\sqrt{s},\rho_A) $. Related works are in progress and will appear elsewhere.

\section*{Acknowledgment}
S.~i.~N. is grateful to fruitful discussions with J.~K.~Ahn (Korea University). This work was supported by the National Research Foundation of Korea (NRF) grants funded by the Korea government (MSIT) (No.~2018R1A5A1025563 and No.~2019R1A2C1005697).


\end{document}